\documentstyle[prl,aps,multicol,epsfig]{revtex}

\begin{document}
\title{Electromagnetically induced transparency in an atom-molecule \\ Bose-Einstein
condensate}
\author{Guang-Ri Jin$^{1}$\cite{PresentAdd,email}, Chul Koo Kim$^{1}$, and Kyun
Nahm$^{2}$}
\address{$^{1}$ Institute of Physics and Applied Physics, Yonsei University, Seoul 120-749, Korea}
\address{$^{2}$ Department of Physics, Yonsei University, Wonju 220-710,Korea}
\date{\today}
\maketitle

\begin{abstract}
We propose a new measurement scheme for the atom-molecule dark
state by using electromagnetically induced transparency (EIT)
technique. Based on a density-matrix formalism, we calculate the
absorption coefficient numerically. The appearance of the EIT dip
in the spectra profile gives clear evidence for the creation of
the dark state in the atom-molecule Bose-Einstein condensate.
\end{abstract}


\begin{multicols}{2}


The phenomenon of dark states is well known in quantum optics and
is based on a superposition of long-lived system eigenstates which
decouples from the light field. The coherent dark states lead to
the phenomenon of electromagnetically induced transparency
\cite{EIT,Scully}. The light-matter coupling associated with the
EIT can be used for the preparation and detection of coherent
matter wave phenomena in ultra-cold quantum gases. For instance,
EIT has been suggested as a probe for the diffusion of the
relative phase in a two-component Bose Einstein condensate (BEC)
\cite{Ruostekoski}. The absorption apectra of the EIT in $\Lambda$
configuration BEC have also been studied both experimentally
\cite{EITBEC1,EITBEC2} and theoretically \cite{EITBEC3,EITBEC4}.
Hau et al., have demonstrated the reduction of the group velocity
of a light pulse to $17$ m/s \cite{Hau}. Using similar
experimental setup, Liu et al., demonstrated the coherent storage
and read out of optical information in the ultracold EIT sample
\cite{Liu}.

Recently, possibility of preparing atom-molecule Bose-Einstein
condensate (AMBEC) has attracted wide attentions
\cite{drum,timm,abee,Wynar}. Condensed bosonic atoms can be
converted to a molecular condensate by using either the
photoassociation process (PA)\cite{drum,Java99,hein} or the
so-called Feshbach resonance method
\cite{timm,abee,Donley,Claussen}. In one-color PA, two colliding
atoms absorb a photon and form an electronically excited molecule.
To overcome the generation of the highly excited molecules,
two-color free-bound-bound mechanism, i.e., stimulated Raman
adiabatic passage was proposed to convert atomic BEC into a
molecular condensate \cite
{Vardi,Jul98,Mackie,Hop01,Drummond,Dam03,HYLing}. The two-color PA
scheme relies on the atom-molecule dark state (AMDS). Besides the
dynamical property of the PA, photoassociative spectroscopy of the
one-color PA \cite{PRL88,PRL91} and the two-color PA
\cite{twocPA1,twocPA2} has been studied intensively in the
ultracold atom-molecule system. Most recently, Winkler et al.,
demonstrated the creation of the AMDS by measuring two-color PA
spectra \cite{AMDS}. Dumke et al., observed similar spectrum
profile in the thermal sodium gas \cite{PDLett}. The PA spectra
measurement is in fact the population of atoms or excited
molecules (i.e., diagonal density matrix elements), which does not
imply the macroscopic coherence between atoms and molecules
\cite{PDLett}.

In this Letter, we propose an absorption imagining study of
electromagnetically induced transparency in a $\Lambda$-type
atom-molecule Bose-Einstein condensate. Our scheme has many
advantages, such as (i) the absorption spectra in multilevel
atomic systems has been well-understood in terms of the dark
states; (ii) the observation time (several $\mu$s) of our scheme
is much shorter than that of Ref. \cite{AMDS} (typical ten ms),
which drastically reduces the influence of the particle losses.
Our study is outlined as follows. First of all, we derive a
density-matrix formalism from a coupled Gross-Pitaevski equations.
Previous results in Ref. \cite{AMDS} are reproduced by calculating
the atomic population. To proceed, we investigate temporal
evolution of the off-diagonal density matrix elements. Our results
show a destructive interference in the AMBEC. We focus on the
absorption coefficient of weak-probe field, and find the EIT dip
in the spectra, which exhibits the creation of the AMDS. Finally,
we present analytically the linear susceptibility of the AMBEC.


We consider a $\Lambda $ configuration atom-molecular BEC system
\cite{AMDS}, where two ground states $\left\vert a\right\rangle $
and $\left\vert g\right\rangle $ are coupled to an intermediate
excited state $\left\vert b\right\rangle $ by two external optical
fields. In the framework of mean-field theory, the dynamics can be
well described by the Gross-Pitaevski equations for the normalized
field amplitudes $a$, $b$, and $g$ \cite
{Vardi,Jul98,Mackie,Hop01,Drummond,Dam03,HYLing} ($\hbar=1$):
\begin{mathletters}
\begin{eqnarray}
\dot{a} &=&i\Omega_{1}a^{\ast}b, \\
\dot{b} &=&\left(i\Delta_1-\gamma _{b}/2\right) b+\frac{i}{2}%
(\Omega _{1}a^{2}+\Omega _{2}g), \\
\dot{g} &=&(i\delta-\gamma _{g}/2)g-\frac{i}{2}\Omega _{2}b,
\label{equ:formel}
\end{eqnarray}%
\end{mathletters}
where $\Omega_j$ with $j=1,2$, are the free-bound and the
bound-bound Rabi frequencies, respectively. Due to
Bose-enhancement, $\Omega_1$ is proportional to
$\sqrt{\varrho_0}$, where $\varrho_0$ is the atomic initial
density. The decay rates $\gamma _{b}$ and $\gamma_{g}$ of the
molecules are included phenomenologically. It was shown that
$\gamma_{g}$ is intensity dependent \cite{AMDS}. The two-photon
detuning is $\delta =\Delta _{1}-\Delta _{2}$, where $\Delta
_{1}=\omega _{1}-(E_{b}-2E_{a})$ and $\Delta _{2}=\omega
_{2}-(E_{b}-E_{g})$. Here, $\omega_j$ with $j=1,2$ are frequencies
of the fields, and $E_\alpha$ with $\alpha=a, b, g$ are the
internal energies of species $\alpha$, respectively. The
atom-atom, atom-molecule interactions contribute mean-field shifts
to the detunings in Eq. (1). For low atomic density, e.g.,
$\varrho_{0}=2\times 10^{14}\,\text{cm}^{-3}$, the shifts are
small and can be neglected.

\vskip -0.7cm
\begin{figure}[htbp]
\begin{center}
\includegraphics[width=7cm, height=8.5cm, angle=0]{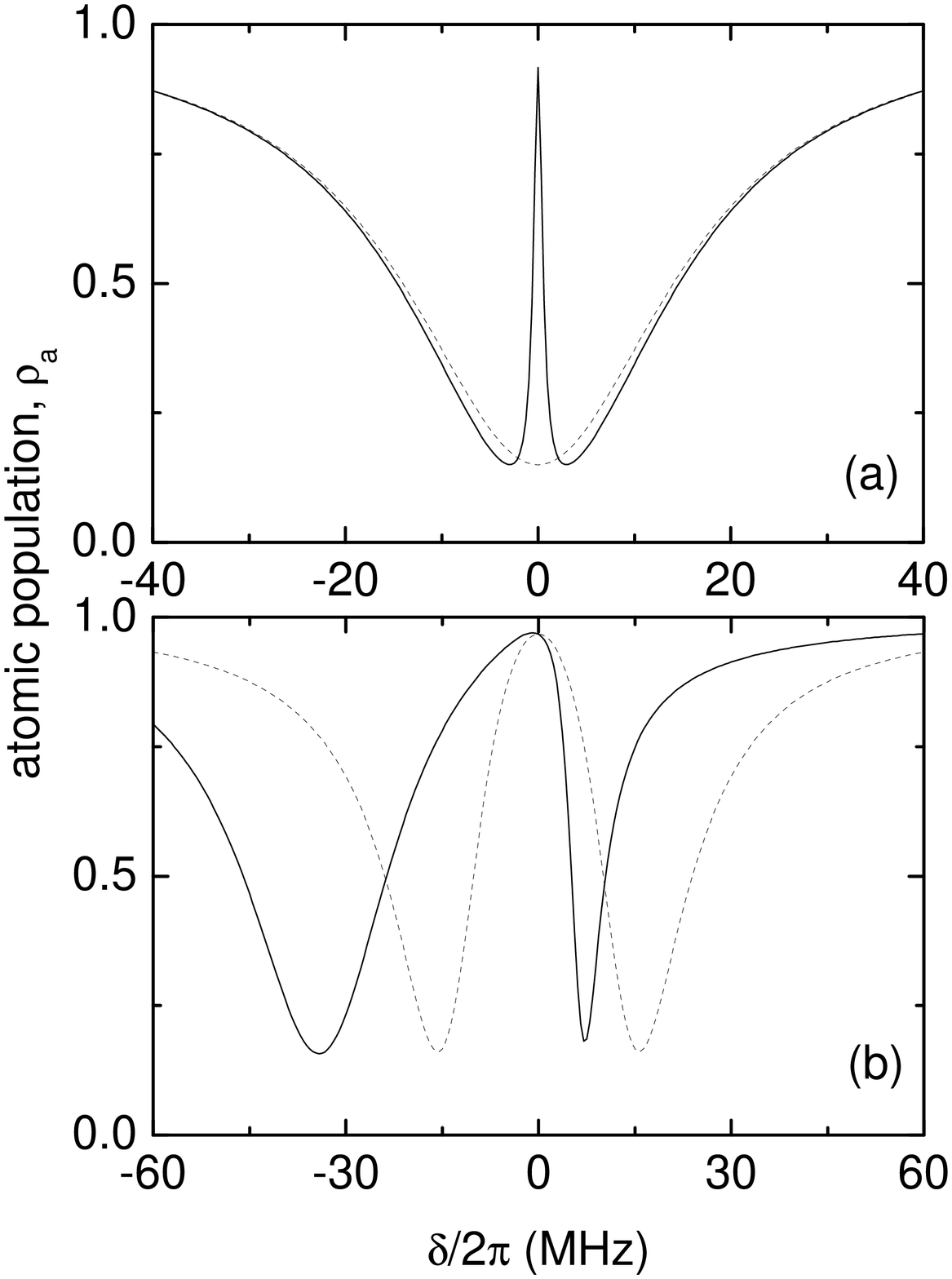}
\end{center}
\caption{Dark resonances in two-color photoassociation spectra for
(a) the low power case $I_{1}=7\text{ W\thinspace cm}^{-2}$, and
$I_{2}=0$ (the dotted line), $0.7$ W\thinspace cm$^{-2}$ (the
solid line) . $\Delta_2=0$; (b) high power case $I_1=80$ W/cm$^2$
and $I_2=20$ W/cm$^2$. The detuning are $\Delta_2=0$ (the dotted
line), $2\pi\times 27$ MHz (the solid line).}
\end{figure}

To investigate atom-molecule coherence, we further introduce
$3\times 3$ dimensional density matrix elements as the following
\cite{density-matrix formalism}: $\rho _{a}=|a|^{2}$, $\rho
_{b}=|b|^{2}$, $\rho _{g}=|g|^{2}$, $\rho _{ba}=b(a^{\ast })^{2}$,
$\rho _{ga}=g(a^{\ast })^{2}$, and $\rho _{bg}=bg^{\ast }$, where
$\rho _{ba}$ ($\rho _{ga}$) describes the photoassociation of two
atoms into one excited (ground) molecule. Within a rotating frame,
the equations of motion for the populations are
\begin{mathletters}
\label{populations}
\begin{eqnarray}
\dot{\rho}_{a} &=&i\Omega _{1}(\rho _{ba}-c.c), \\
\dot{\rho}_{g} &=&-\gamma _{g}\rho _{g}+i\frac{\Omega
_{2}}{2}(\rho
_{bg}-c.c), \\
\dot{\rho}_{b} &=&-\gamma _{b}\rho _{b}-i\frac{\Omega
_{1}}{2}(\rho _{ba}-c.c)-i\frac{\Omega _{2}}{2}(\rho _{bg}-c.c).
\end{eqnarray}%
In the absence of molecular decaying, i.e., $\gamma _{b}=\gamma
_{g}=0$, the total atom number is conserved and $\rho_{a}+2(\rho
_{b}+\rho _{g})=1$. The coherence part density matrix elements
obey
\end{mathletters}
\begin{mathletters}
\label{coherences}
\begin{eqnarray}
\dot{\rho}_{ga} &=&(i\delta -\frac{\gamma _{g}}{2})\rho
_{ga}-2i\Omega _{1}\rho_{a}\rho _{bg}^{\ast }+i\frac{\Omega
_{2}}{2}\rho _{ba},\\
\dot{\rho}_{ba} &=&(i\Delta _{1}-\frac{\gamma _{b}}{2})\rho
_{ba}+i\frac{\Omega _{1}}{2}(\rho _{a}^2-4\rho _{a}\rho
_{b})+i\frac{\Omega _{2}}{2}\rho
_{ga}, \\
\dot{\rho}_{bg} &=&(i\Delta _{2}-\frac{\Gamma}{2})\rho
_{bg}+i\frac{\Omega _{1}}{2}\rho _{ga}^{\ast }+i\frac{\Omega
_{2}}{2}(\rho _{g}-\rho _{b}),
\end{eqnarray}
\end{mathletters}
where $\Gamma =(\gamma _{b}+\gamma _{g})$. Unlike the
$\Lambda$-type atom system, the second terms of Eqs. (3a) and (3b)
are nonlinear due to the free-bound transition.

We solve the above coupled differential equations
(\ref{populations}) and (\ref{coherences}) numerically using a
fourth-order Runge-Kutta routine. Following Ref. \cite{AMDS}, we
study a dilute $^{87}$Rb BEC with a peak density
$\varrho_{0}=2\times 10^{14}\,\text{cm}^{-3}$, then the parameters
are $\gamma _{b}=2\pi \times 13\,\text{ MHz}$, $\Omega
_{1}/\sqrt{I_1}=2\pi \times 8\,\text{kHz}/( \text{W\thinspace
cm}^{-2})^{1/2}$, and $\Omega _{2}/\sqrt{I_{2}}=2\pi \times
7\text{ MHz}/(\text{W\thinspace cm}^{-2})^{1/2}$. The decay rate
of the ground-state molecules is $\gamma _{g}=2\pi \times
6\,\text{kHz}/(\text{ W\thinspace cm}^{-2})I_{1}+\gamma
_{\text{bg}}$, where the background decay rate $\gamma
_{\text{bg}}\approx 2\pi \times 1\,\text{kHz}$ at the peak density
$\varrho_{0}$.

In Fig. 1(a), we consider the low-power field with $I_{1}=7 \text{
W\thinspace cm}^{-2}$, and $I_{2}=0$ (the dotted line), $0.7$
W\thinspace cm$^{-2}$ (the solid line), respectively. Our results
agree with the experimental results as expected \cite{AMDS}. In
one-color photoassociation, the spectrum for a dipole-allowed
transition is a Lorentzian with a width set by the optical
transition linewidth. The presence of the coupling laser field
results in the dark resonance appeared near the two-photon
resonant point $\delta=0$. The linewidth is much narrower than the
natural linewidth of the excited molecules $\gamma _{b}$. In Fig.
1(b), we calculate $\rho_a(\delta)$ for relatively high laser
power $I_{1}=80$ W/cm$^{2}$ and $I_{2}=20$ W/cm$^{2}$ with the
detunings $\Delta _{2}=0$ (the dotted line), $2\pi\times 27$ MHz
(the solid line). The results show relative broadened dark
resonance lines \cite{AMDS}, and nonzero $\Delta _{2}$ results in
asymmetric spectrum, as shown by the solid line in Fig. 1(b).


\vskip -0.7cm
\begin{figure}[htbp]
\begin{center}
\includegraphics[width=7cm, height=8.5cm, angle=0]{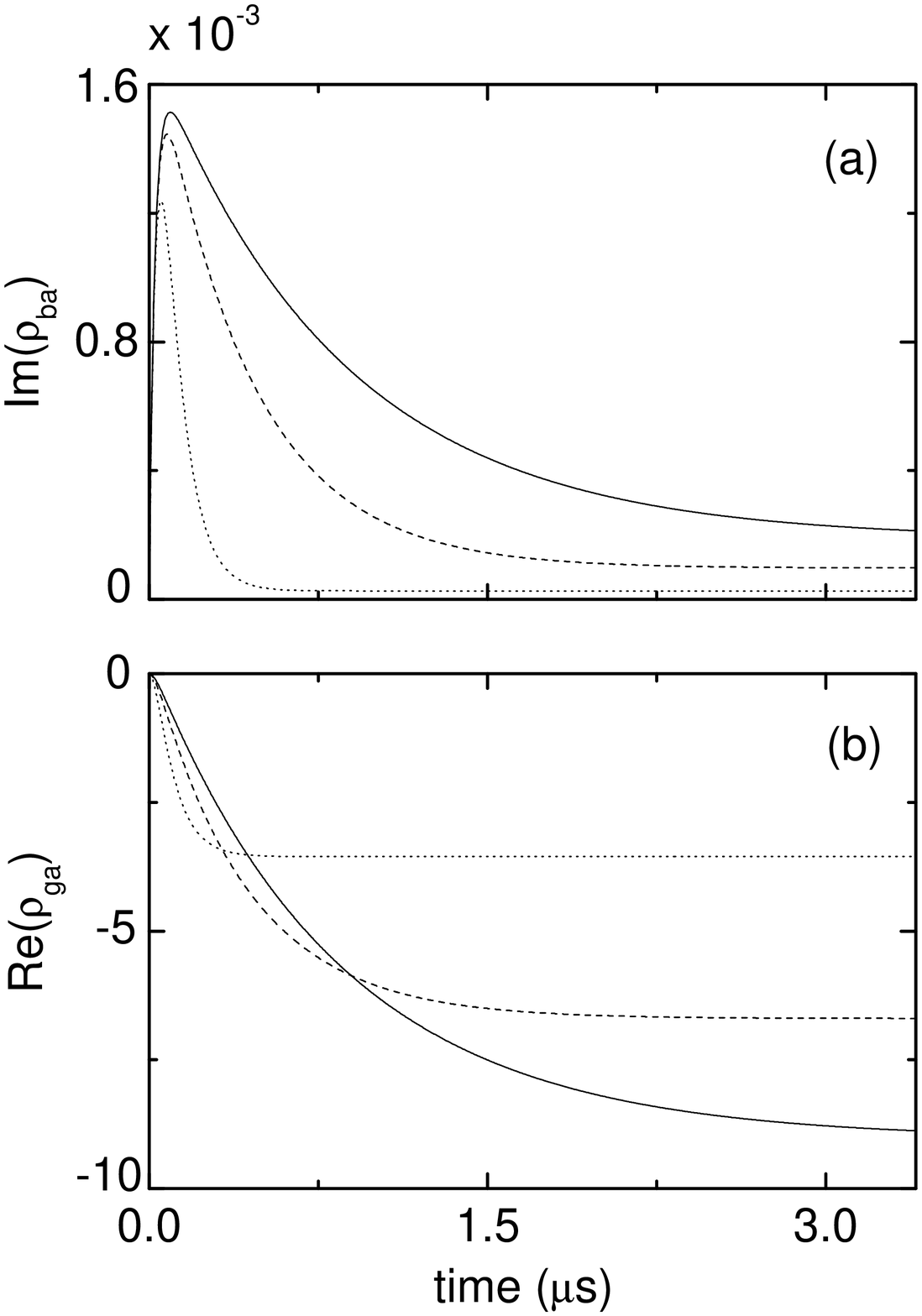}
\end{center}
\caption{Time evolution of Im($\rho_{ba}$) and Re($\rho_{ga}$) for
$I_{1}=7\text{ W\thinspace cm}^{-2}$, and $I_{2}=0.7$ W\thinspace
cm$^{-2}$ (the dotted lines), $0.18$ W\thinspace cm$^{-2}$ (the
dashed lines), and $I_{1}/80$ (the solid lines).
$\Delta_1=\Delta_2=0$.}
\end{figure}

The diagonal matrix element $\rho_a$ does not imply a coherent
superposition of atoms and molecules \cite{PDLett}. To investigate
the macroscopic coherence, we propose to measure the probe-field
absorption coefficient, which is proportional to Im($\rho _{ba}$).
Fig. 2 shows temporal evolution of Im($\rho _{ba}$) and Re($\rho
_{ga}$) for low-power case. We find that Im($\rho _{ba}$) grows
quickly then decays to the steady-state values around $3.4 \mu s$.
The steady-state value of Im($\rho _{ba}$) depends on the coupling
field Rabi frequency $\Omega_2$. The larger $\Omega_2$ is, the
smaller the value is. Fig. 2(b) shows that Re($\rho_{ga}$) is
always negative during total evolution, which implies destructive
interference between the two competitive transitions.

\vskip -0.7cm
\begin{figure}[htbp]
\begin{center}
\includegraphics[width=7cm, height=8.5cm, angle=0]{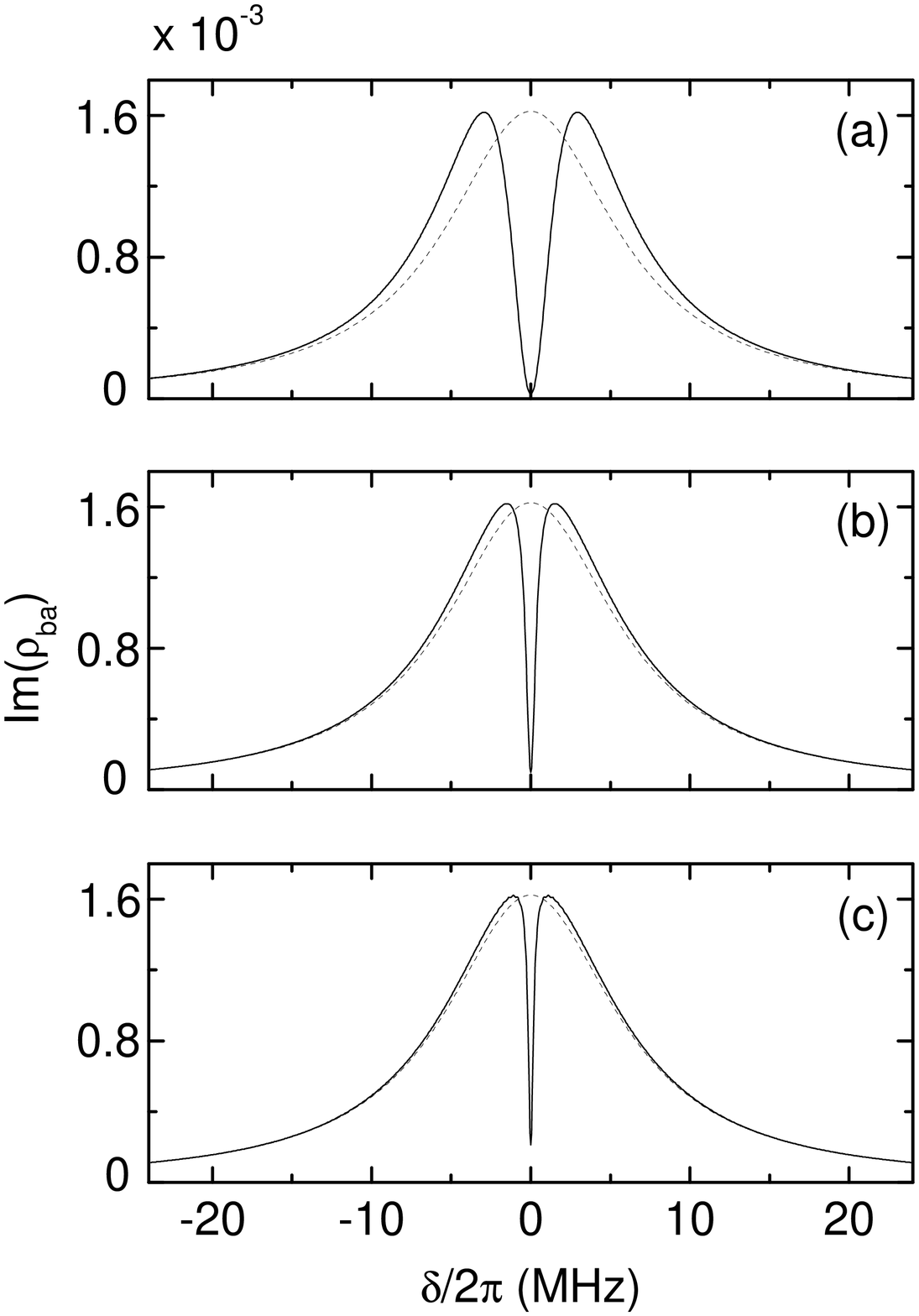}
\end{center}
\caption{Absorption spectra, Im($\rho_{ba}$) for $I_{1}=7\text{
W\thinspace cm}^{-2}$, (a) $I_{2}=0.7$ W\thinspace cm$^{-2}$, (b)
$0.18$ W\thinspace cm$^{-2}$, (c) $I_{1}/80$. $\Delta_2=0$. The
dotted line is the case of $I_{2}=0$.}
\end{figure}

We calculate the absorption spectra of the probe field from Eqs.
(\ref{populations}) and (\ref{coherences}) for $I_{1}=7\text{
W\thinspace cm}^{-2}$. In the absence of the coupling field,
Lorentzian shape of Im($\rho _{ba}$) corresponds to the resonant
absorption in the two-level system (see dotted lines in Fig. 3).
For $\Omega_2\neq 0$, the absorption spectrum exhibits an EIT dip
near the resonant point $\delta=0$. The depth and width of the dip
are dependent on the coupling field Rabi frequency $\Omega_2$. For
a relatively larger Rabi frequency, e.g., $I_2=0.7$ W\thinspace
cm$^{-2}$, the medium becomes almost transparent. With the
decrease of $I_2$, the molecular losses become dominant, and the
transparency is no longer perfect.

Time-evolution of Im($\rho _{ba}$) and Re($\rho _{ga}$) for the
high-power cases with $I_{1}=80\text{ W\thinspace cm}^{-2}$ and
$I_{2}=20\text{ W\thinspace cm}^{-2}$ exhibit damped oscillations,
which tend to steady-state values at about $1\mu s$. Again
Re($\rho _{ga}$)$<0$ during time-evolution indicates the
destructive interference in the high-power regions. In Fig. 4 (a),
we calculate the absorption spectrum for $\Delta_2=0$. The
absorption profile shows the signatures of an Autler-Townes
doublet. The separation of two peaks are about $2\pi\times 30$
MHz, which agrees with the magnitude of Rabi frequency $\Omega_2$.
For nonzero $\Delta_2$, Im($\rho _{ba}$) oscillates irregularly
with time evolution, and the spectra become asymmetric and exhibit
a board and narrow peak, as shown in Fig. 4(b) and 4(c).

According to Ref. \cite{Drummond}, the Rabi frequencies take the
forms of $\Omega_1=d_1{\cal E}_1{\cal I}^{(1)}\sqrt{\varrho_0/2}$
and $\Omega_2=d_2{\cal E}_2{\cal I}^{(2)}$, where $d_j$ ($j=1,2$)
are mean dipole matrix moments, ${\cal E}_j$ the slowly varying
field amplitudes, and ${\cal I}^{(j)}$ the Franck-Condon overlap
integrals, respectively. A factor of $\sqrt{2}$ is introduced in
$\Omega_1$ to consistent with the notation in Ref.
\cite{Drummond}. The linear susceptibility is
$\chi^{(1)}=\frac{2\varrho_0d_1^*}{\epsilon_0{\cal
E}_1}\rho^{(1)}_{ba}$, where $\rho^{(1)}_{ba}$ is the first-order
solution of $\rho_{ba}$. Within the weak-probe regime, i.e.,
$\Omega_1\ll \Omega_2$, $\gamma_b$, the absorption of the AMBEC
can be well-described by the first-order steady-state solution
$\rho_{ba}^{(1)}=i\Omega_1/(2F)$, where $F=\gamma_b/2-i\Delta
_{1}+\frac{\Omega _{2}^{2}/4}{\gamma_g/2-i\delta}$ \cite{MXiao}.
Combining the above discussions, we obtain the linear
susceptibility of the AMBEC system
\begin{equation}
\chi^{(1)}=\frac{i\varrho_0^{3/2}|d_1|^2{\cal
I}^{(1)}}{\sqrt{2}\epsilon_0 F}.
\end{equation}
For $^{87}$Rb BEC, ${\cal I}^{(1)}\simeq10^{-11}$cm$^{3/2}$ and
${\cal I}^{(2)}\simeq0.1$ \cite{hein,Drummond}. Considering the
density $\varrho_0=2\times10^{14}\text{cm}^{-3}$ and the probe
field intensity $I_1=7\text{W}\text{cm}^{-2}$, we obtain ${\cal
E}_1=\sqrt{2I_1/(c\epsilon_0)}=7.264\times10^{3}$V/m.
Consequently, we can calculate the mean dipole matrix moment
$d_1=19.3\times10^{-30}$Cm. It was shown by Ref. \cite{PRL91}, the
dipole matrix moment of the free-bound transition to the excited
molecular state is relatively smaller than the atomic dipole.

\vskip -0.7cm
\begin{figure}[htbp]
\begin{center}
\includegraphics[width=7cm, height=8.5cm, angle=0]{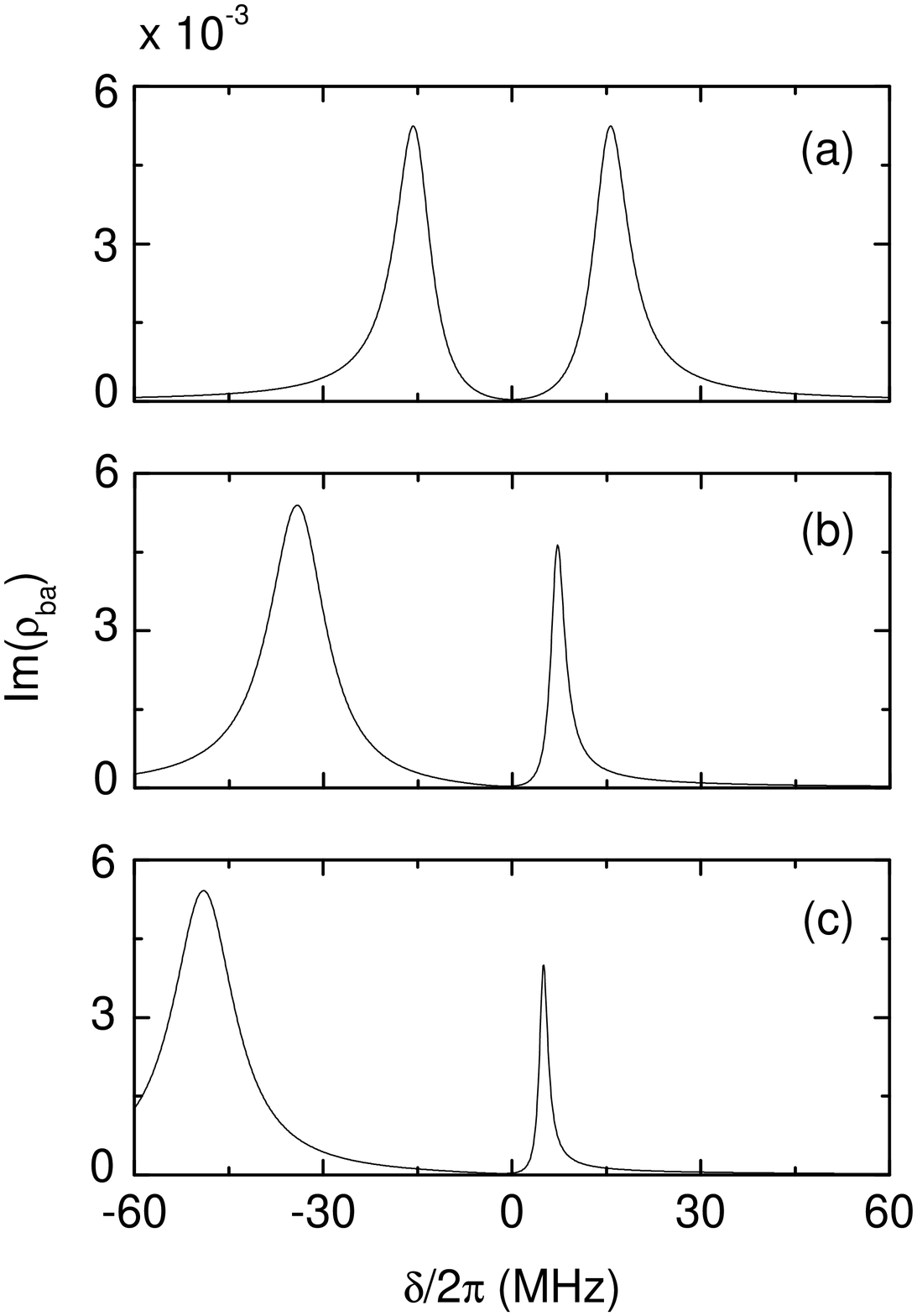}
\end{center}
\caption{Absorption spectra, Im($\rho_{ba}$) for $I_{1}=80\text{
W\thinspace cm}^{-2}$, $I_{2}=20\text{ W\thinspace cm}^{-2}$. (a)
$\Delta_2=0$, (b) $\Delta_2=2\pi\times 27$ (MHz), and (c)
$\Delta_2=2\pi\times 44$ (MHz).}
\end{figure}

In summary, we have proposed the possibility of observing the EIT
in the AMBEC system. From the introduced density matrix formalism,
we study how the transparency of a probe beam of free-bound
transition can be controlled via a coupling light beam resonant
with the bound-bound transition. All of our results are consistent
with that of Ref. \cite{AMDS}. We show that in weak-probe limit,
the absorption profile can be well-described by the first-order
steady-state solution of the matrix element. We present analytical
expression of the linear susceptibility $\chi ^{(1)}$, and
calculate electric dipole moment, $d_1$, for the free-bound
transition. Finally, we would like to emphasize that, compared
with the previous photoassociative spectroscopy  \cite{AMDS}, the
measurement of probe-field absorption has its advantage to give
direct coherence information of the AMBEC dark state, and the
observation time of our scheme is much shorter so that the
particle dissipation effects can be reduced significantly.

This work is supported in part by the BK21 and by KOSEF through
Center for Strongly Correlated Materials Research, SNU (2005).

\end{multicols}
\end{document}